\documentclass[aps,pra,amsmath,amssymb,preprint]{revtex4-1}
\usepackage{graphicx}
\usepackage{bm}
\usepackage{mathptmx}
\usepackage{epstopdf}
\usepackage{booktabs}
\usepackage{multirow}
\usepackage{hhline}
\usepackage{subfigure}
\begin{document}
\newcommand{\ms}{m_*}
\newcommand{\bor}{{\bf r}}
\newcommand{\mm}{{\bf m}}
\newcommand{\rdot}{\dot{\bf r}}
\newcommand{\rddot}{\dot\dot{\bf r}}
\newcommand{\es}{\bf{E^s_\sigma}}
\title{Spin torques and anomalous velocity in spin textures induced by fast electron injection from topological ferromagnets: The role of gauge fields}
\author{Satadeep Bhattacharjee$^{1}$}
\email{s.bhattacharjee@ikst.res.in}
\affiliation{Indo-Korea Science and Technology Center (IKST), Bangalore, India}
\author{Seung-Cheol Lee$^2$}
\affiliation{Electronic Materials Research Center, Korea Institute of Science $\&$ Technology, Korea}
\begin{abstract} 
A new method for analyzing magnetization dynamics in spin textures under the influence of fast electron injection from topological ferromagnetic sources such as Dirac half metals has been proposed. These electrons, traveling at a velocity $v$ with a non-negligible value of $v/c$ (where c is the speed of light), generate a non-equilibrium magnetization density in the spin-texture region, which is related to an electric dipole moment via relativistic interactions. When this resulting dipole moment interacts with gauge fields in the spin-texture region, an effective field is created that produces spin torques. These torques, like spin-orbit torques that occur when electrons are injected from a heavy metal into a ferromagnet, can display both damping-like and anti-damping-like properties. Finally, we demonstrate that such an interaction between the dipole moment and the gauge field introduces an anomalous velocity that can contribute to transverse electrical conductivity in the spin texture in a way comparable to the topological Hall effect.
\end{abstract}
\maketitle
\section{Introduction}
The impact of relativistic effects on current-induced magnetization switching in ferromagnets is a fascinating area of research, attracting interest from both practical and academic viewpoints \cite{interface4}. Such effects are often explored in bilayer nanowires that consist of a ferromagnetic layer and a non-magnetic layer with strong spin-orbit interaction \cite{int1,interface3}. The non-magnetic layer conducts a charge current that generates a spin current perpendicular to the interface, which exerts a spin transfer torque on the magnetization of the ferromagnetic layer. The magnetization dynamics that result seem to involve torques that can be divided into damping-like and field-like components. Although limited, there have been some studies exploring the magnetic order of ferromagnets with non-trivial textures in the context of current-induced magnetization switching \cite{int2}. Non-collinear magnetic textures have recently been at the forefront of the field of spintronics due to the exciting applications and opportunities they offer \cite{text1,text2}. Magnetic films having non-collinear magnetic configurations such as skyrmions, vortices, and so on are thought to have superior qualities to collinear magnets and could be employed in data storage and processing systems. Magnetic textures such as skyrmions have been proposed for the racetrack types of memories where they can move via injection of spin polarized currents \cite{RT}. The interplay between electron transport and magnetic textures affects electron transport in metallic magnets. Magnetic textures affect electron transport through the Berry phase, they accumulate on the electron wave function, resulting in modulation of the electron spin by the localized spins. This Berry phase effect give rise to the so called \textit{spin electromagnetic fields} that act on the majority and minority spin states of the itinerant electrons and alter electron transport \cite{volovik1987}.

The damping parameter is a crucial factor for spintronic systems that involve magnetization switching \cite{damp1,damp2,damp3}. Magnetization dynamics are often explored using the Landau-Lifshitz-Gilbert (LLG) equation, in which the damping term is used in a phenomenological approach to describe how the spin-angular momentum is dissipated from the magnetic system. In most real applications, an optimal value of the damping is needed. On one hand, a higher damping is desirable for a quicker switching rate, while on the other hand, for reducing the current needed for switching (low power devices, less heating), lower damping is preferable. Therefore, for a variety of applications, an appropriate Gilbert damping is required \cite{bhat}.

In this work, we investigate the tuning of damping in a magnetic texture due to the injection of spin-polarized electrons moving at very high speeds.  In recent years there is an increased focus on magnetic topological materials because of their possible various applications~\cite{MTM}. This work represents a ferromagnetic topological material, Dirac half-metal (DHM) as a carrier injector with very high Fermi velocity and spin polarization. Typical prototypes of such DHM are, for example, YN$_2$, Rhombohedral Lanthanum Manganites or recently proposed triangular magnets ~\cite{YN2,manganite,tria}. 

This study proposes that the injection of spin-polarized carriers into a spin texture region can produce an additional effect beyond the usual s-d type interaction with the background spin texture, which typically generates gauge electric and magnetic fields. This additional effect arises from the high drift velocity of the carriers, which induces an electric dipole moment that can couple with the gauge electric field, resulting in an additional torque that significantly impacts the dynamics of the spin texture. It is worth noting that in a material with a linear band structure like graphene, the drift velocity can be on the order of Fermi velocity \cite{drift1}. For instance, Shishir \textit{et al.} have demonstrated that the drift velocity in graphene can be as high as $4.5 \times 10^5$ m/s \cite{shishir}. Quantum topological materials can exhibit ultra-low effective masses and very large Fermi velocities, making them promising candidates for carrier injectors with high drift velocity or mobility. Ultimately, we demonstrate that such interaction between the electric dipole moment and the gauge fields can result in an anomalous carrier velocity that further influences the transport behavior in the spin texture region.  

The interplay between the gauge electric field and dipoles yields two notable effects, as revealed by our study. Firstly, it leads to a significant modification of the damping parameter when fast-moving spin-polarized electrons are injected from the DHM layer. Secondly, it results in anomalous carrier velocity, which can impact the transport behavior in the spin texture region. These findings have important implications for the design and optimization of spintronic devices, as both effects play critical roles in ensuring efficient and reliable magnetization switching. 
\section{Effective field and the torque due to the interaction between the gauge field and the dipole moment}
\begin{figure}
\centering
\includegraphics[scale=0.4]{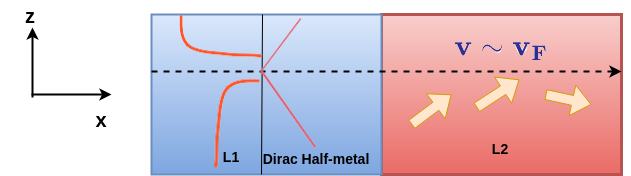}
\caption{Schematic figure showing the geometry of the layer structure and the co-ordinate system. The red curves on the left side represent the density of states for the spin-up and spin-down electrons in the DHM region. The thick yellow arrows on the right side represent the magnetic texture.}
\label{schem}
\end{figure}

In this section, we introduce the spin torque exerted by the magnetization density of the conduction electron on the  spin texture via an effective field which generated through an interaction between the electric dipole moment of the conduction electrons moving at relativistic speed and the gauge electric field. In Fig.\ref{schem}, we show a schematic of the possible geometry for studying the above mentioned effects. The spin polarized electrons are injected from the left from the region L1  to the region L2  by an electric field. As mentioned above, region L1 could be a DHM while region L2 is the magnetic texture. It can be noted that the spin electromagnetic 
fields as behave like classical objects which merely depend on the magnetization of the layer L2 containing the chiral moments described by a vector  $\bf M(r)$ given by, ${{\bf M}({\bf r},t)}=(sin\theta cos\phi,sin\theta sin\phi,cos\theta)$.
The spin electromagnetic fields are generated through the interaction between the itinerant  electrons with the spin texture can be described in terms of the so-called sd-Hamiltonian~\cite{gauge1,gauge2,gauge3} in the region L2 given by,
\begin{equation}
{\hat H}=\frac{\hat{\bf p}^2}{2m_*}-J_{sd}{\bf \sigma}\cdot{\bf M(\bf r,t)}
\label{sd}
\end{equation}
Here ${\bf M(\bf r,t)}$ describes the magnetic texture that varies in time and space. $m_*$ is the effective mass of the conduction electron. $J_{sd}$ describes the coupling between the magnetization texture and the spin of the conduction electrons described by the Pauli spin matrices ${\bf \sigma}$. $\hat{\bf p}$ is the electron momentum. 
The nature of this coupling is ferromagnetic i,e $J_{sd} > 0$. By performing a SU(2) unitary transformation to the above Hamiltonian using the operator, $U=e^{\dfrac{i\theta }{2}\sigma _{y}}e^{\dfrac{i\phi }{2}\sigma _{z}}$ which rotates the magnetization vector ${\bf M}$ along a fixed quantization axis to produce the so-called spin electromagnetic fields. The i$^{th}$ component of the spin electromagnetic fields that appear are  given by, by~\cite{volovik1987,kohno2014theoretical},
\begin{equation}
\begin{split}
E^s_i=\pm \frac{\hbar}{2e}{\bf M}\cdot(\dot{\bf M}\times \nabla_i \bf M)\\ 
B^s_i=\pm \frac{\hbar}{2e}\epsilon_{ijk}{\bf M}\cdot(\nabla_j \bf M\times \nabla_k \bf M)
\end{split}
\label{fields}
\end{equation}
Here the sign $\pm$ depends on the type of carrier (either spin up or down).
The emergent magnetic and electric fields mentioned above do not directly couple to the magnetic texture itself. They act only on the conduction electrons. However, they can indirectly affect the magnetic texture  via their coupling to the conduction electrons.  One has to note here that the emergent electric and magnetic fields shown above are in their simplest form: They are formulated under the assumption that the carriers in the region L2 follow the magnetic texture adiabatically. Let us consider the coupling between the gauge electric field, ${\bf E}^s$ with conduction electrons in terms of the electric dipole moment of such conduction electrons. The electric dipole moment of the electrons associated with magnetization density with a velocity ${\bf v}$ is given by~\cite{book,vek,djg},
\begin{equation}
{\bf d}= \frac{1}{c^2}(\bf v\times {\bf m})
\label{paradox}
\end{equation}
Here c is the velocity of the light. The magnetization density associated with conduction electrons is denoted by $\mathbf{m}$ and is obtained by averaging out the quantum mechanical degrees of freedom as $\mathbf{m} = \langle \boldsymbol{\sigma} \rangle$. This quantity can be divided into two parts as $\mathbf{m} = \mathbf{m}_0 + \delta \mathbf{m}$, where $\mathbf{m}_0$ represents the equilibrium magnetization density of the conduction electrons in the L2 region without any injected carriers, and $\delta \mathbf{m}$ represents the perturbation to the equilibrium magnetization density due to carrier injection from L1. Usually, the equilibrium part does not produce any torque as it aligns with ${\bf M}$. The non-equilibrium part, however, is usually misaligned with ${\bf M}$ and can provide a torque~\cite{R1}.

In this work, we are not interested in such direct torque that is described in the literature~\cite{R1}. Instead, we focus on an indirect process that arises due to the interaction between ${\bf d}$ and ${\bf E}^s$, which will be presented below.

It is important to note, however, that visualization of the non-equilibrium magnetization density requires the use of a half-metallic source, as we have not considered the spin-orbit interaction in the L2 region. Another perspective on the same phenomenon is that the spin texture (L2) exhibits strong spin-orbit interaction, while the topological material (L1) is non-magnetic. 
The particular form of the electric dipole due to the moving magnetic dipole shown in the Eq. \ref{paradox} was discussed in details~\cite{djg} in the context of so-called \textit{Mansuripur paradox} which involves a discrepancy in observing a torque exerted by a point charge on a magnetic dipole in two different frames of reference, violating the principle of relativity~\cite{paradox}. The paradox was resolved in terms of the so called \textit{hidden momentum} associated with the magnetic dipole if one considers it to be an Ampere dipole. Also, the dipole moment in this form is significant in the relativistic description of spin precession in the presence of a static homogeneous electromagnetic field, as demonstrated in the Thomas-Bergmann-Michel-Telegadi model~\cite{TBMT} and Frenkel \textit{et al.}'s seminal work on the subject~\cite{p2}.

Moving on, the energy associated with the interaction between this dipole moment and the \textit{spin electric field} caused by non-trivial magnetic order in the region L2 is given by, 
\begin{equation}
{\cal E} = -{\bf E^s}\cdot{\bf d}
  =-\sum_i\frac{\hbar}{2e}{\bf M}\cdot({\dot{\bf M}}\times \nabla_i {\bf M}) d_i
\label{interaction}
\end{equation}
$d_i$ is the i$^{th}$ component of the dipole moment. The magnetic texture in the region L2 can experience an effective magnetic field due to the above interaction of the conduction electrons with the gauge electric field,
\begin{equation}
 {\bf H}^{g}=-\frac{\partial {\cal E}}{\partial {\bf M}}
=\frac{\hbar}{2e}\sum_i(\dot{\bf M}\times \nabla_i {\bf M})d_i
\label{field}
\end{equation}


Therefore the additional torque acting on the localized moments in the region L2 is given by,
\begin{equation}
{\bf T^g}= {\bf M} \times {\bf H}^{g}
=\frac{\hbar}{2e} \sum_i d_i{\bf M}\times(\dot{\bf M}\times \nabla_i {\bf M})=\frac{\hbar}{2e} \sum_i d_i{\bf M}\times(\dot{\bf M}\times \nabla_i {\bf M})
\label{torque}
\end{equation}
 Now adding this torque to the LLG equation we get,
\begin{equation}
\begin{split}
\dot{ \bf M}=&\gamma {\bf M} \times {\bf H_{eff}} +\alpha {\bf M} \times \dot {\bf M} +\gamma{\bf T^g}\\
=&\gamma {\bf M} \times {\bf H_{eff}} +\alpha {\bf M} \times \dot {\bf M} +{\bf M} \times (\mathcal D\cdot \dot {\bf M})
\end{split}
\label{LLG}
\end{equation}
Here $\alpha$ is the intrinsic damping in the region L2 which is mainly due to the spin-orbit interaction of the material.  The above equation gives tensorial damping factor given by $\mathcal{D}_{k,l}= \frac{\hbar\gamma}{2eM_s} \sum_i d_i M_k({\bf M}\times \nabla_i {\bf M})_l$. $M_s$ being the saturation magnetization.
\\

\begin{figure}
\centering
\includegraphics[scale=0.8]{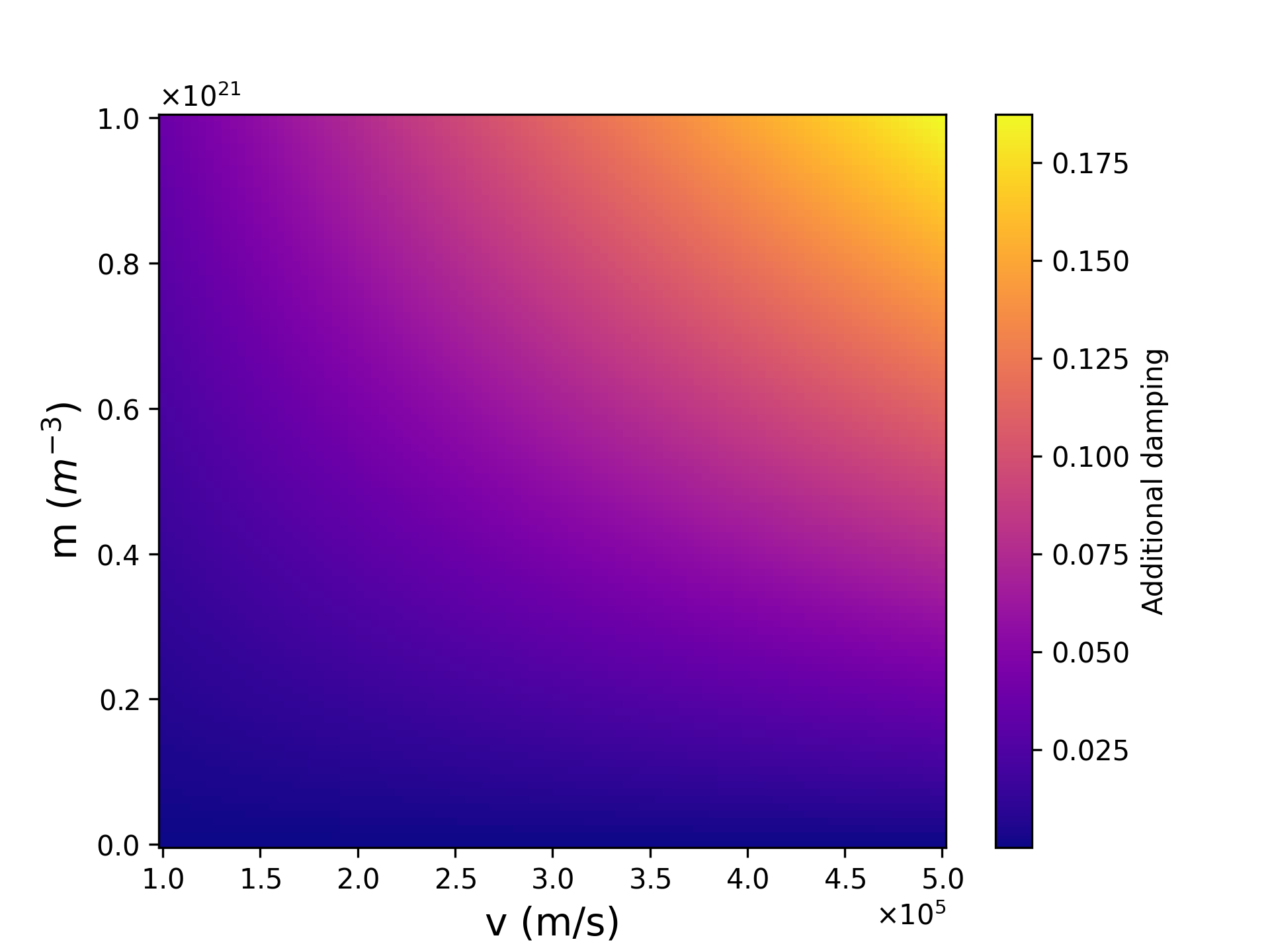}
\caption{Heatmap of the additional damping for the realistic materials parameters.}
\label{add}
\end{figure}

Here we have used the fact that $(\dot{\bf M}\times \nabla_i {\bf M})=[(\dot{\bf M}\times \nabla_i {\bf M})\cdot {\bf M}]{\bf M}$. If we consider the geometry shown in the Fig.\ref{schem}, the velocity of the electrons have only x-component, i,e ${\bf v}=(v,0,0)$. We get from the above
$\mathcal{D}_{k,l}= \frac{\hbar\gamma}{2eM_sc^2}vM_k[m_y({\bf M}\times \nabla_z {\bf M})_l-m_z({\bf M}\times \nabla_y {\bf M})_l]$. The new damping factor thus depends directly on the velocity of the incident electrons. The total damping is therefore,
\begin{equation}
    \mathcal{D}^T_{k,l}=\delta_{k,l}\alpha+ \frac{\hbar\gamma}{2eM_sc^2}vM_k[m_y({\bf M}\times \nabla_z {\bf M})_l-m_z({\bf M}\times \nabla_y {\bf M})_l]
\end{equation}
Usually, $\alpha$ varies from 0.001 to 0.01 in most of the magnetic materials. Fig.\ref{add} displays the behavior of additional damping with various realistic material parameters. Specifically, we use the following values: The reduced Planck constant $\hbar = 1.0545718 \times 10^{-34}  \text{J}\cdot\text{s}$, the gyromagnetic ratio $\gamma = 1.7608597 \times 10^{11}  \text{rad}/(\text{s}\cdot\text{T})$, the elementary charge $e = 1.60217662 \times 10^{-19} \text{C}$, the magnetization saturation $M_s = 1.72 \times 10^{6}  \text{A}/\text{m}$, and the speed of light $c = 3 \times 10^8  \text{m}/\text{s}$. We consider a range of velocities between $10^5  \text{m}/\text{s}$ and $10^6 \text{m}/\text{s}$ for the injected carriers which is consistent with recently observed Fermi velocities in the Dirac half metals proposed for the so called \textit{high speed spintronics}~\cite{spped}. The magnetization density of the carriers are defined as $m=n_\uparrow-n_\downarrow$, while $n_\uparrow(n_\downarrow)$ is the electron density in the spin-texture region for the spin up and down electrons. In the figure we have varied the magnetization density $m$ within the range $10^{18}\text{m$^{-3}$}$ to $10^{21}\text{m$^{-3}$}$. The figure shows how the additional damping term evolves with crucial parameters such as the velocity and the concentration of the injected carriers.
The LLG equation above takes the form in terms of the total damping tensor as,
\begin{equation}
\dot{ \bf M}=\gamma {\bf M} \times {\bf H_{eff}} + {\bf M} \times (\mathcal D^T\cdot \dot {\bf M})
\end{equation}
It can be seen that the nature of this new damping term depends on the direction of the spin-polarization of the incident electrons. In the case, when the incident electrons are spin-polarized along the z-direction $[{\bf m}=(0,0,m_z)]$, this additional damping actually acts opposite to the intrinsic damping $\alpha$ and lowers the effective damping in the system which is given by, $\mathcal{D}^T_{k,l}=\delta_{k,l}\alpha-\frac{\gamma}{2M_sc^2}vm_zM_k({\bf M}\times \nabla_y {\bf M})_l$. Thus this torque is a kind of anti-damping  torque that tries to amplify the processional motion. On the other hand, if the incident electrons are polarized along y-direction $[{\bf m}=(0,m_y,0)]$, one can easily see that  it basically enhances the total damping in the system. Therefore the above interactions give both damping-like and anti-damping like contributions depending on the direction of the polarization of the incident electrons. Such damping and anti-damping torques are usually discussed in the context of current induced magnetization switching in a ferromagnet in the proximity of a heavy nonmagnetic metal. In one scenario, due to the spin Hall Effect~\cite{SOV}, a spin current is generated in the heavy metal~\cite{interface1,interface2,interface3} which when enters in the ferromagnetic region exerts both damping and anti-damping like torques. However, these torques are determined by the bulk electronic states or more precisely the bulk spin-Hall angle in the heavy non-magnetic metal. In another scenario,  where both damping and anti-damping torques are generated by the surface states and usually described in terms of a Rashba model~\cite{R1,R2,R3}.  There are some similarity between the nature of the torque we obtain with the latter scenario~\cite{R1,R2,R3} such as in both cases the torque depends on the velocity of the incident electrons and their spin polarization.  However, the distinct difference with our cases is that we need a magnetic texture in the ferromagnetic region to account for the gauge fields. The torques derived by the above-mentioned works~\cite{interface3,R1,R3} are independent of the magnetization gradient whereas in our case the torques depend on the magnetization gradient $\nabla_i {\bf M}$ as can be seen from the Eq.\ref{torque}. Another important difference comes from the role of the exchange coupling $J_{sd}$. In the above studies, the torque is due to the direct coupling between the magnetization density ${\bf m}$ and the ferromagnetic background ${\bf M}$ either via $\frac{J_{sd}}{\hbar}{\bf M}\times {\bf m}$ or $\frac{J_{sd}}{\hbar}{\bf M}\times ({\bf m}\times {\bf M} )$ type of interaction, while we do not have any direct coupling between
${\bf m}$ and ${\bf M}$ here, rather an indirect coupling is present which is given by $\frac{1}{c^2}[{\bf v}\times {\bf E^s}({\bf M},{\bf \nabla M})]\cdot{\bf m}$. The role of $J_{sd}$ here is limited to generating ${\bf E}^{s}$.
Finally, the source of the non-equilibrium magnetization in the ferromagnetic layer is not the spin-orbit interaction in the proximal layer, rather it is obtained via the spin injection from a DHM.
Apart from that,  the gauge electric field is a central quantity here and to observe this effect the incident spin density should have a non-negligible $v/c$ ratio. 

Our study clearly demonstrates that the damping in ferromagnetic textures can be effectively controlled using ferromagnetic topological materials as a source. By injecting spin-polarized currents, it is possible to engineer the damping in magnetic textures, leading to more efficient manipulation and potentially achieving lower damping. This has significant implications for the future of magnetic memory technology, as it could result in faster and more reliable operation of memory devices. In skyrmion-based MRAM, our study suggests that lower damping, achieved through the use of ferromagnetic topological sources, may lead to more stable skyrmions that can be manipulated with lower current densities. The above study therefore clearly demonstrates that using a ferromagnetic topological source can play a very important role in the future, as it is possible to engineer the damping in magnetic textures via injecting spin-polarized current. By using ferromagnetic topological sources and spin-polarized currents, it may be possible to achieve lower damping and more efficient manipulation of magnetic textures in memory devices, leading to faster and more reliable operation. Therefore, this study clearly demonstrates the potential of using ferromagnetic topological sources to play a very important role in the future of magnetic memory technology. Particularly, in skyrmion-based MRAM, damping can affect the stability and motion of the skyrmions, which are topologically protected magnetic textures. Lower damping can lead to more stable skyrmions that can be manipulated with lower current densities.
\section{Torque and the effective field from the continuity equation of the magnetization density}

In this section, we will demonstrate that we can derive the same expression for the effective field and the torque using the so-called continuity equation of the magnetization density.
Including the above interaction, the total Hamiltonian can be written  as,
\begin{equation}
{\hat H}=\frac{\hat{\bf p}^2}{2m_*}-J_{sd}{\bf \sigma}\cdot{\bf M}+\dfrac{1}{m^*c^{2}}\left( {\bf E}^{s}\times {\bf p}\right) \cdot {\bf m}
\label{Again}
\end{equation}
Here as already mentioned, ${\bf m}=<\sigma>$ and ${\bf p}=<\hat{\bf p}>$ is the classical momentum for the electron. Using the continuity equation for the magnetization density of the carriers,
\begin{equation}
\dfrac{d{\bf m}}{dt}=-{\bf \nabla}{\bf j_{s}}-\dfrac{J_{sd}}{\hbar }{\bf M}\times {\bf m}+\dfrac{1}{\hbar m_*c^{2}}\left( {\bf E}^{s}\times{\bf p}\right) \times {\bf m}
\label{cont}
\end{equation}

Here ${\bf j_{s}}$ is the spin current tensor. Considering a steady state and the incident carriers are uniformly polarized, which should in principle be a good approximation for the source such as Dirac half-metal, ${\bf \nabla}{\bf j_{s}}=0$, we can find the torque again,

\begin{equation}
{\bf T}=J_{sd}{\bf M}\times {\bf m}=\dfrac{1}{m_*c^{2}}\left( {\bf E}^{s}\times{\bf p}\right) \times {\bf m}
\label{torque2}
\end{equation}
We can extract the effective field from the above equation,
\begin{equation}
\begin{split}
{\bf T}=&{\bf E}_{s}\times {\bf d}\\
=&\dfrac{\hbar }{2e}\sum _{i}{\bf M}\cdot\left({\dot{\bf M}}\times \nabla_i{\bf M}\right)\hat{\bf e}_i\times {\bf d}\\
=&\dfrac{\hbar }{2e}{\bf M}\times \sum _{i}\left({\dot{\bf M}} \times \nabla_{i}{\bf M}\right)d_i\\
=&{\bf M}\times {\bf H}_{eff}
\end{split}
\end{equation}

The effective field therefore is given by ${\bf H}_{eff}=\dfrac{\hbar }{2e} \sum _{i}\left(\dot{\bf M} \times \nabla_{i}{\bf M}\right)d_i$ which has the same form
of ${\bf H}^g$ as obtained in Eq.\ref{field}.
\section{New anomalous velocity and transverse conductivity}
In this final section, we show that an additional anomalous velocity contribution can be accounted for as  a result of the interaction between the electric dipole moment of such relativistic magnetic dipoles and the spin gauge fields that arise due to the Berry phase effects. This effect  gives rise to a transverse electrical conductivity that depends on the non-zero scalar triple product of three vectors as seen in the usual topological Hall effect.
To describe the transport properties, we start with the following Hamiltonian,
\begin{equation}
{\mathcal H}=\dfrac{{\bf p}^{2}}{2\ms}+V\left( {\bf r}\right)+{\bf E}^{s}_{\sigma}\cdot {\bf d}=\dfrac{{\bf p}^{2}}{2\ms}+V\left( {\bf r}\right)+\frac{1}{\ms c^2}{\bf p}\cdot (\mm\times \es)
\end{equation}
Here $V\left( {\bf r}\right)$ describes the potential due to various sources other than the lattice which has been taken care of in terms of the effective mass $\ms$. We also use the subscript for the electric field in order to differentiate the spin-up and spin-down carriers.
From the above Hamiltonian, we can obtain the velocity and forces given by
\begin{equation}
\begin{split}
\rdot&=\frac{\partial {\mathcal H}}{\partial{\bf p}}=\frac{p}{\ms}+\frac{1}{\ms c^2}(\mm\times \es)\\
\dot{\bf p}&=-\frac{\partial {\mathcal H}}{\partial{\bf r}}=-\dfrac{\partial V\left( {\bf r}\right) }{\partial {\bf r}}-\dfrac{1}{\ms c^{2}}\left( {\bf p}\times \mm \right) \left( \dfrac{\partial }{\partial {\bf r}}\cdot\es\right)
\end{split}
\end{equation}
By simplifying the above two equations we obtain Newton's equation of motion given by,
\begin{equation}
\dot{\bf p}=-\dfrac{\partial V\left( {\bf r}\right) }{\partial {\bf r}}+{\bf F}_\sigma
\label{Newton}
\end{equation}
Here ${\bf F}_{\sigma}$ is the spin dependent force ${\bf F}_{\sigma}=-\dfrac{1}{c^{2}}\rdot\times \left[ \dfrac{\partial }{\partial {\bf r}}\times \left( \mm \times {\bf E}_{\sigma }^{s}\right) \right]$. If we compare this force with the standard Lorentz force given by, ${\bf F}=e(\rdot\times {\bf B})$, we obtain a spin-dependent magnetic field of ${\bf B}_{\sigma }=\dfrac{1}{ec^{2}}\dfrac{\partial }{\partial {\bf r}}\times \left( \mm\times {\bf E}_{\sigma }^{s}\right)$ and a vector potential ~${\bf A}_{\sigma }=\dfrac{1}{ec^{2}} \left( \mm\times {\bf E}_{\sigma }^{s}\right)$. If we consider that the spin texture varies only in xy-plane, then using the semi-classical approach used by Bruno \textit{et. al}~\cite{Bruno}, we can derive the Hall conductivity as,

\begin{equation}
\sigma _{xy}=\sigma _{xx}\left( \dfrac{e <B_{z}>\tau }{\ms}\right)
            =z_s\dfrac{\sigma _{xx}\tau m_z }{\ms c^2}\int \left[\partial_xE^s_x+\partial_yE^s_y\right]dxdy
\label{div}
\end{equation}
In the above equation, we have dropped the spin index $\sigma$ before the magnetic field, rather we use the variable $z_s$ to distinguish the majority and the minority carriers. As the form of the integral suggests a 
planer geometry is used to calculate the average magnetic field.
$\tau$ is the relaxation time and $\sigma_{xx}$ is the usual transverse conductivity due to the applied electric field which drives the carriers from the topological materials to the spin texture. Here we have assumed ${\bf m}=(0,0,m_z)$.  The above contribution of Hall conductivity in addition to the usual  topological one, $\sigma^{top} _{xy}=\sigma _{xx}\int {\bf M}\cdot \left( \partial _{x}{\bf M}\times \partial_y {\bf M}\right) dxdy$ which is resulting from the magnetic field shown in the Eq.\ref{fields}. The term on the right-hand side tells that the additional contribution is equal to the divergence of the spin-electric field in the plane of the sample. 
\par
By combining Eq.(\ref{fields}) and Eq. (\ref{div}), we now introduce a new quantity that is similar to the topological charge in the case of the well-known topological Hall effect given by,
\begin{equation}
Q=\int {\bf M}\cdot\left[(\partial_x{\dot{\bf M}}\times \partial_x{\bf M})+(\partial_y{\dot{\bf M}}\times \partial_y{\bf M})\right]dxdy
\end{equation}
The above quantity, therefore, represents a scalar triple product of three spins spins
${\bf M}$, $\partial_k{\dot{\bf M}}$ and $\partial_k{\bf M}$ where $k=x,y$. The quantity $Q$ can be thought of as very similar to the skyrmion number~\cite{n1,n2} in the literature which also depends on the scalar triple product. However, it can be seen that in the present case, Q is non-zero only when there is a time evolution of the magnetic texture. 
\par
This study presents a new type of topological Hall effect that arises for carriers influenced by spatially varying magnetization. While the mechanism shares similarities with the original topological Hall effect, (in that it is also linked to the Berry phase acquired by an electron moving in smoothly varying magnetization), the specific details differ significantly. According to the semiclassical model that we proposed here, the Lorentz force acting on carriers is not due to the gauge magnetic field resulting from the Berry phase effect. Instead, it is due to the interaction between the gauge electric field and the electric dipole moment associated with the carriers. This dipole moment is only significant when carriers are injected from topological materials, such as Dirac half metals.

Therefore, this work offers an alternative approach to access spin chirality in novel magnetic materials, allowing the distinction between different magnetic textures such as domain walls and skyrmions. These distinctions are highly sought after in the spintronics field as they can support the development of more advanced spintronics devices.
\section{conclusions}
To summarize, we studied magnetization dynamics in a chiral ferromagnet proximal to a magnetic topological material, such as Dirac half-metal, under the influence of an electric field injecting fast carriers into the chiral magnet. Our findings revealed the emergence of additional damping terms resulting from the interaction between the electric dipole moment of the fast magnetization density and the gauge electric field in the chiral magnet region. The type of damping term observed depended on the spin polarization direction of the incoming magnetization density and could be either damping-like or anti-damping-like. Additionally, we observed that the interaction between the electric dipole moment associated with the fast magnetization density and gauge electric field resulted in a new anomalous velocity that could affect the transverse electronic transport of the spin texture.
\section{Acknowledement}
This work was supported by the Korea Institute of Science and Technology, GKP (Global Knowledge Platform, Grant number 2V6760) project of the Ministry of Science, ICT and Future Planning.
\bibliography{Damp}
\bibliographystyle{apsrev4-1}
\end{document}